%% file: main.tex
\documentclass[sigconf,nonacm]{acmart}

\usepackage{silence}
\setcopyright{none}
\renewcommand\footnotetextcopyrightpermission[1]{}
\makeatletter
\gdef\@acmBooktitle{}%
\gdef\@acmConference{}%
\gdef\@acmYear{}%
\gdef\@acmPrice{}%
\gdef\@acmDOI{}%
\gdef\@acmISBN{}%
\makeatother
\usepackage{algorithm}
\usepackage[noend]{algpseudocode}
\usepackage{paralist}
\usepackage{epsfig}   %%  for postscript graphics files
\usepackage{mathtools}
\usepackage{float}
\usepackage{graphicx}
\usepackage[export]{adjustbox}

\usepackage{siunitx}
\usepackage{tabularx}
\usepackage{subcaption}
\usepackage{listings,multicol}
\usepackage{setspace}
\usepackage{comment}
\usepackage{multirow}
\usepackage{tcolorbox}
\usepackage[linewidth=1pt]{mdframed}
\usepackage{colortbl}   %color, 
\usepackage{textcomp}
\usepackage{enumitem}
\usepackage{epstopdf}
\usepackage{todonotes}
\usepackage[font={small}]{caption}

\usepackage{wrapfig,lipsum,booktabs}
\usepackage{subcaption}
\usepackage{booktabs}
\usepackage{array}

\definecolor{blue(pigment)}{rgb}{0.2, 0.2, 0.6}
\usepackage{bbding}       %%tick, cross
\usepackage{nccmath}      %% for large (m, k)
\usepackage{dblfloatfix}  %% for position of table
\usepackage{pifont}       %%tick, cross
\usepackage[T1]{fontenc}   %% for position of table
\usepackage[utf8]{inputenx}   % for position of table
\usepackage{caption}            %% space between table and caption
\usepackage{optidef}
\usepackage[toc]{appendix}
\usepackage{tikz}   %% for drawing figures in latex

\title{Precision Switching Schedule for Efficient Control Implementations}
\author{Debarpita Banerjee}
\email{debarpita2023_r@isical.ac.in}
\affiliation{
  \institution{Indian Statistical Institute}
  \city{Kolkata}
  \country{India}
}

\author{Debasmita Lohar}
\email{debloh@itu.dk}
\affiliation{
  \institution{IT University of Copenhagen}
  \city{Copenhagen}
  \country{Denmark}
}

\author{Sumana Ghosh}
\email{sumana@isical.ac.in}
\affiliation{
  \institution{Indian Statistical Institute}
  \city{Kolkata}
  \country{India}
}
\begin{document}
%%%=====================abstract==================
\input{abstract}

\maketitle

%%%============== sections ===============
\input{sections/introduction}

\input{sections/background}
\input{sections/technique}

\input{sections/evaluation}
\input{sections/rq2}

\input{sections/rq3}

\input{sections/related}

\input{sections/conclusion}

%========================================	
 %========================================
%\bibliographystyle{ACM-Reference-Format}
\bibliographystyle{abbrv}
\bibliography{ref}

\end{document}

%% file: abstract.tex
\begin{abstract}
Modern cyber-physical systems, such as automotive control, rely on feedback controllers that regulate the system towards desired a setpoint. In practice, however, the controller must also be scheduled efficiently on resource-constrained processors, where the choice of numerical precision for controller implementation directly affects both control quality and computational cost. This trade-off is critical: higher precision improves control performance but increases runtime, while lower precision executes faster in the processor but may degrade overall system performance.

\sloppy
In this work, we propose the first approach for a \emph{precision-switching schedule}, where the controller switches between different floating-point precisions to balance control performance and enhance computational efficiency. We formulate this problem as a multi-objective optimization, expressed as a Mixed-Integer Quadratic Program (MIQP) with sound linearizations and error bounds that capture roundoff effects from different precision implementations. Our method efficiently computes a switching schedule that ensures the system output remains within a specified reference band. Through experimental evaluation on standard benchmark control systems, we demonstrate that switching between 32-bit and 16-bit floating-point implementations offers an average runtime reduction of 26.5\% compared to 32-bit execution and a 27.6\% improvement in control performance over 16-bit execution, while maintaining near-optimal overall performance.

\end{abstract}
\keywords{Plant-control systems, real-time scheduling, floating-point arithmetic, control performance requirements}

%% file: sections/introduction.tex
\section{Introduction}
\label{sec:introduction}

Modern cyber-physical systems, such as those in automotive control, avionics, and robotics, widely employ feedback controllers that continuously adjust control inputs to steer the system output toward desired reference values. 
These controllers are often executed as real-time control tasks on the processor, which must run periodically and complete within strict timing requirements or deadlines to preserve closed-loop performance and responsiveness. 
Therefore, designing effective scheduling strategies for controllers remains an important problem.

There exists a large body of work on real-time scheduling of controllers ~\cite{scheduling_single-energy,maggio2020stability,hobbs2022safety} and co-scheduling of multiple control loops~\cite{scheduling_control_CPS,scheduling_CPS,ghosh_17,banerjee_FMSS,ghosh_19}, which explores various aspects like control performance and stability under timing constraints~\cite{ghosh_17,DeepCAS_performance_scheduling,scheduling_stability_performance,vreman2021stability,online2018stability,scheduling_codesign_stability}. 
However, these works do not consider how the controllers' implementations affect their behavior. In practice, controllers are deployed on hardware platforms that support finite-precision (floating- or fixed-point) arithmetic; hence, the precision used during execution directly affects both control performance and computational cost. 
This aspect --- how the choice of precision influences the system's control performance requirements while scheduling the controller on the processor --- has been largely overlooked in the control scheduling literature.

\sloppy
This choice of precision plays a crucial role in determining both control performance and computational cost. Low-precision arithmetic offers faster computation; however, these computations can accumulate roundoff errors, leading to degraded accuracy and potentially unstable closed-loop systems. 
In contrast, high-precision arithmetic improves accuracy as well as performance but has increased runtime. 
Balancing this fundamental trade-off is thus essential for achieving correct and efficient control.

Recently, several works have considered the role of numerical precision in control-theoretic contexts such as robust controller synthesis~\cite{precision2019memory}, safety analysis~\cite{harikishan2024precision}, and mixed-precision quantization of neural-network controllers~\cite{lohar2023sound}. 
However, none have explored the effects of precision on the system's performance requirements when scheduling the controller on the processor.

To bridge this gap, we introduce, to the best of our knowledge, the first \textit{precision-aware scheduling framework} for controllers. Our approach efficiently computes optimized \textit{switching schedules} that enable the controller to switch between different floating-point precision levels at runtime, maintaining control performance requirements while substantially reducing overall execution time.

We formulate this scheduling problem as a multi-objective \textit{mixed-integer quadratic program} (MIQP) that jointly optimizes control performance and execution time. 
The objective is to maximize control performance, quantified by minimizing the linear quadratic cost (LQR)~\cite{ogata2020controlBook}, while minimizing total computation time. The optimization is performed subject to the constraint of meeting the system's settling time requirement, which serves as another performance metric that characterizes the responsiveness of the physical process regulated by the feedback controller. 
Our formulation explicitly incorporates roundoff errors introduced by finite-precision arithmetic to ensure the soundness of the resulting switching schedule. 

A direct formulation of this problem leads to non-linear constraints, which are computationally intractable.
To address this, we also introduce a combination of exact linearizations and sound overapproximations of roundoff errors that relax the problem into one with linear constraints, making it efficiently solvable by state-of-the-art MIQP solvers. 

Solving this optimization problem yields an optimal precision-switching schedule of the controller that guarantees that the system output remains within a specified error margin around the desired reference. 
Although our current formulation focuses on switching between two precision levels, high and low, it can be naturally extended to multiple precision options.

We have implemented our approach in a prototype tool that automatically formulates the optimization problem based on the control system's dynamics, reference and performance metrics (rise time, peak time, and settling time, and employs state-of-the-art solvers to generate optimized schedules. This tool will be released as open source upon acceptance.

%%%%%%%%%%%%%%%%%%%%%%%%% MOTIVATIONAL EXAMPLE %%%%%%%%%%%%%%%%%%%%%%%
\paragraph{\textbf{Running Example}}
\label{subsec:intro_example}
Let us consider a third-order \emph{cruise control} (\texttt{CC}) system, a common benchmark in automotive applications. 
The discrete-time dynamics of the system are given by $x[k+1] = A x[k] + B u[k], y[k] = C x[k]$, where

{
\[
A = \begin{bmatrix}
0 & 0.01 & 4.99e{\text -}5\\
{\text -}3.02e{\text -}4 & 0.99 & 0.01\\
{\text -}0.06 & {\text -}0.05 & 0.99
\end{bmatrix}, \,
B = \begin{bmatrix} 4.13e{{\text -}7} \\ 1.24e{{\text -}4} \\ 0.02 \end{bmatrix},\,
C = \begin{bmatrix} 1 & 0 & 0 \end{bmatrix}.
\]
}
$x[k]$, $u[k]$, and $y[k]$ denote the state, control input, and output at the $k$-th sampling instant, respectively, and $K$ represents the controller gain used to compute $u[k] = Kx[k-1]$.
The goal of this controller is to ensure that the system output $y[k]$ follows the desired reference over time.

\begin{figure}[t]
    \centering   
    \includegraphics[scale=0.43]{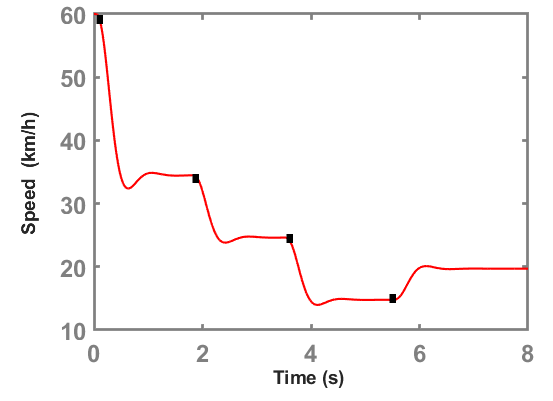}   
    \caption{Cruise control (CC) simulation}
    \label{fig:CC_trajectory_intro}
\end{figure}

~\autoref{fig:CC_trajectory_intro} shows the response  of \texttt{CC} when the reference speed is changed to $\SI{35}{km/h}$, $\SI{25}{km/h}$, $\SI{15}{km/h}$, and $\SI{20}{km/h}$ at $\SI{0}{\second}$  $\SI{1.8}{\second}$, $\SI{3.5}{\second}$, and $\SI{5.5}{\second}$, respectively.

We consider that the controller can switch between two floating-point precision options: 16-bit (FP16) and 32-bit (FP32), and the control performance is measured using the LQR cost, where a lower value indicates better performance.

The system is simulated for $\SI{8}{\second}$ (800 samples, with a $\SI{10}{\milli\second}$ sampling period) on an NVIDIA GPU supporting both precision options. 
Executing the controller entirely in FP16 results in degraded control performance, resulting in an \textit{NaN} LQR cost due to severe roundoff accumulation. 
Using FP32, the controller achieves an LQR cost of $341106$, but the runtime increases by 57.14\% ($\SI{0.21}{\milli\second}$ for FP32 versus $\SI{0.09}{\milli\second}$ for FP16). 
This contrast highlights the inherent trade-off between precision and computational efficiency, along with the impact on the control performance, motivating the need for a precision-aware schedule of controllers.

Our proposed precision-switching scheduling technique automatically determines when to execute the controller in FP16 or FP32 during runtime. For this CC system, the generated optimal schedule uses FP32 for the first $210$ samples, FP16 for the next $34$, FP32 for the next $109$, and so on, with FP32 again used for the last $154$ samples. The resulting LQR cost is $341041$, almost identical to FP32 execution, while reducing the runtime to $\SI{0.17}{\milli\second}$, a $19.05\%$ improvement over full FP32 execution.

%%%%%%%%%%%%%%%%%%%%%%%%%  CONTRIBUTIONS %%%%%%%%%%%%%%%%%%%%%%%
\paragraph{\textbf{Contributions}}
In summary, this paper makes the following contributions.
\begin{enumerate}
    \item We introduce the first \textit{precision-switching scheduling} technique for controllers that determines when and how to schedule the controller at different floating-point precision levels to optimize control performance and efficiency.

\item We formulate the scheduling problem as a multi-objective mixed-integer quadratic program (MIQP) that jointly optimizes control performance and execution time while meeting the settling time requirement. The formulation explicitly incorporates roundoff errors arising from finite-precision arithmetic and employs sound over-approximations and linearizations to make the problem efficiently solvable.

\item We evaluate the proposed method on multiple automotive control benchmarks. Experimental results demonstrate that switching between 32-bit and 16-bit floating-point controller implementations achieves an average runtime reduction of 26.5\% compared to full 32-bit execution and a 27.6\% improvement in control performance over 16-bit execution, while maintaining near-optimal control performance.
\end{enumerate}

%% file: sections/background.tex
\section{Background}
\label{sec:background}
This section provides a brief overview of a real-time plant-control closed-loop system, its properties, and performance metrics (\autoref{subsec:plant-control}), as well as the fundamentals of floating-point arithmetic (\autoref{subsec:floats}), both of which will be used throughout the paper.

\subsection{Real-Time Closed-Loop Control Systems}
\label{subsec:plant-control}
A plant-controller system is a closed-loop continuous system where a dynamical process (the \textit{plant}) is regulated by a stabilizing feedback controller. The controller observes the plant's output and computes the control input to maintain a desired behavior.

Although such linear time-invariant (LTI) systems are typically described in continuous time, in embedded systems the controller operates periodically on a processor with a fixed \textit{sampling period}. The dynamics are therefore represented in discrete time as:

\begin{equation}
\label{eq:discrete-state-eqtn}
x[k+1] = A x[k] + B u[k], \quad y[k] = C x[k],
\end{equation}
where \( x[k], u[k], y[k] \) are the discrete-time state, control input, and output at the \( k \)-th sampling instant, termed \textit{sample}, and \( A, B, C \) are the corresponding discrete-time system matrices.

The controller must compute its output within a \textit{deadline}, which is typically equal to the sampling period. Under the \textit{logical execution time} paradigm~\cite{maggio2020stability}, the control input is computed based on the previously sampled state and applied exactly at the deadline. The dynamics of the discrete-time controller is follows.

\begin{equation}
\label{eq:control input}
u[k] = K x[k-1]
\end{equation}
Here, \( K \) is the feedback gain matrix. In this work, we design the gain matrix using the \textit{Linear Quadratic Regulator} (LQR) method~\cite{ogata2020controlBook} on \textit{static controllers}, where the control action depends solely on the latest sampled state.

\paragraph{Control Performance Requirements}
Control implementation must ensure  the underlying \textit{performance requirements} of the system. 
We use the \textit{settling time}~\cite{ogata2020controlBook} as one of the performance metrics, defined as the time required for the system output to reach and remain within a specified \textit{output reference band}, i.e., the output reference value~$\pm$~a predefined small error margin (typically \(1\%\)--\(5\%\) of the reference), even under sudden step inputs. The plant exhibits a \textit{transient response} while approaching the reference band after a step input and reaches a \textit{steady state} once it remains within this band. A shorter settling time indicates a faster responsiveness for a stable dynamical system, making it a key metric for evaluating plant performance. In hard real-time systems, meeting all deadlines together with a finite settling time ensures asymptotic stability.

To evaluate control performance, specifically in terms of control effort, we use the standard \textit{LQR cost} function:
\begin{equation}
\label{eq:lqr_cost}
LQR\ cost = \sum_{k=0}^{\infty} \left( x^\mathsf{T}[k] Q x[k] + u^\mathsf{T}[k] R u[k] \right)
\end{equation}

Here, \( Q \succcurlyeq 0 \) and \( R \succ 0 \) are symmetric weight matrices that penalize state deviation and control effort, respectively~\cite{astrom97}. Minimizing the cost improves control efficiency and resource usage.

Other performance metrics include \emph{peak time}, which is defined as the time required to reach the first overshoot after a step input, and \emph{rise time}, defined as the duration for the output to rise to 90\% of the reference~\cite{ogata2020controlBook}.
Smaller values for these metrics indicate a faster, more responsive system.

\paragraph{Real-Time Scheduling of a Control Task} 
A digital controller is typically implemented as a \emph{software control task} on the underlying computing platform. This task is subject to real-time constraints and must complete its execution within its deadline to ensure correct system operation. The objective is to determine a scheduling strategy for the control task that guarantees both control performance requirements --- minimizing the LQR cost and meeting the settling time requirement.

\subsection{Floating-Point Arithmetic}\label{subsec:floats}
Floating-point numbers are the standard representation of real numbers in modern hardware due to their efficiency and widespread support. Their finite precision, however, introduces roundoff errors that propagate through computations and may amplify or dampen depending on the operations and input ranges. 
To model these errors, analyses typically follow the IEEE~754 standard~\cite{ieee-standard}:

\[
x \circ_{\text{fl}} y = (x \circ y)(1 + e) + d, \qquad |e| \leq \varepsilon_m,\ |d| \leq \delta_m
\]

where $\circ \in {+, -, *, /}$, and $\varepsilon_m$ and $\delta_m$ denote relative and absolute error bounds for normal and subnormal values. Common formats include \emph{half} (16-bit), \emph{single} (32-bit), and \emph{double} (64-bit) precision, with $\varepsilon_m$ values of $2^{-11}$, $2^{-24}$, and $2^{-53}$, and $\delta_m$ values of $2^{-24}$, $2^{-150}$, and $2^{-1075}$. This model assumes rounding to the nearest and excludes overflow, which analyses usually prove cannot occur.

\paragraph{Worst-Case Error Analysis} 
Most roundoff error analyses bound the worst-case absolute error between real-valued computations $f(x)$ and their floating-point counterparts $\tilde{f}(\tilde{x})$ over bounded domains: $
\max_{x \in [a,b]} |f(x) - \tilde{f}(\tilde{x})|$.
Representative methods include dataflow analyses such as interval and affine arithmetic, used in tools like Daisy~\cite{daisy} and Fluctuat~\cite{fluctuat}, and global optimization–based methods employed by FPTaylor~\cite{fptaylor}, PRECiSA~\cite{precisa}, and Satire~\cite{satire}.

%% file: sections/technique.tex
\section{Precision-Aware Scheduling}
\label{sec:method}

The goal of this work is to develop a scheduling strategy for controllers operating at multiple floating-point precisions that maintains the control performance requirements while reducing execution time.
While always using high precision ensures good control performance, it also leads to high runtime. Conversely, low-precision execution runs faster but may degrade control quality. The key challenge lies in determining an effective trade-off for when each precision should be used.

Given the controller configuration, system dynamics, and performance criteria, along with the time instants of applying step inputs and their corresponding reference outputs, our objective is to find an optimized \emph{switching schedule}, if one exists, that determines when different precisions should be applied. This schedule allocates the low- and high-precision controller implementations on the processor to minimize total execution time without compromising the performance.

We formulate this scheduling as a multi-objective optimization problem that jointly maximizes control performance and minimizes total execution time under step inputs. The optimization is subject to the constraint that, once the system has settled after a step input, the output remains in a specified reference band.

Additionally, roundoff errors arising from floating-point arithmetic must be accounted for when computing the switching schedule. To ensure this, we introduce additional constraints in the optimization problem that incorporate the roundoff errors of the plant states, control inputs, and outputs.

However, directly encoding the switching decisions and roundoff errors within the optimization problem introduces nonlinear constraints, which are typically non-convex and often intractable in practice. To make the problem tractable, we use sound over-approximations by performing linearizations where necessary and by precomputing sound bounds on the roundoff errors. The discrete-time LQR cost function used in the objective introduces quadratic terms, resulting in a Mixed-Integer Quadratic Program (MIQP). Despite being NP-hard in general, our problems remain convex and can be efficiently solved using state-of-the-art solvers~\cite{gurobi,cplex_code}.

In the subsequent sections, we present the optimization problem formulation in detail and describe each step of the proposed method.

\subsection{The MIQP Problem}
\label{sub:generalized-optimization-problem}

The goal of our MIQP formulation is to determine a switching schedule composed of a low-precision execution sample and a high-precision one. For brevity, we refer to them as \emph{lo} and \emph{hi}, respectively.
We focus on scheduling a single controller that switches between these two floating-point precision levels.

The system is assumed to start with a step input, requiring an initial \emph{hi}-precision sample for better control performance during the transient response. Subsequent samples may switch between \emph{hi} and \emph{lo}-precision based on performance requirements and the system's behavior, at runtime. We also assume that the bounds on state and input variables are known or can be conservatively estimated. While this work considers two precisions, our formulation can be straightforwardly extended to support multiple precision levels and other data types.

Our formulation has two objectives: (1) maximize control performance, which in this context corresponds to minimizing the LQR cost (as defined in \autoref{eq:lqr_cost}) and (2) minimize the total execution time across the \emph{hi} and \emph{lo} precision executions. The optimization is subject to the constraint that the system meets its settling time requirement, i.e., the output remains within the reference band (as defined in~\autoref{subsec:plant-control}). 

We first introduce the relevant variables used in the MIQP problem in~\autoref{tab:optimization_parameters}, where variables specific to sample and precision are referenced with the subscripts $i$ and $p$, respectively.

\begin{table}[t]
\small
\centering
    \renewcommand{\arraystretch}{1.2}
\begin{tabular}{l|l|l}
\toprule
\textbf{type} & \textbf{variable} & \textbf{description} \\ \midrule
user-defined & $A, B, C$ & system matrices \\
& $K$ & controller gain \\
& $Q, R$ & LQR weight matrices \\
& $h$ & sampling period \\
& $x_0, u_0$ & initial state and control input \\
& $T_s$ & settling time\\
& $\delta$ & error margin around reference \\  
& $T$ & total simulation time \\
& $\{t_j\}_{j=1}^r$ & step input instants \\
& $\{\gamma_j\}_{j=1}^r$ & output reference in $[t_j, t_{j+1}]$ \\
& $t_p$ & runtime per sample in $p$ \\ \midrule
precomputed & $\mu$ & no. of switching intervals \\
& $\mathcal{I}=\{[L_\beta,U_\beta]\}_{\beta=1}^\mu$ & switching intervals \\
& $e_p$ & roundoff error in $p$ \\ \midrule
decision var. & $sw_i$ & precision type\\ \midrule
other var. & $x_i, u_i, y_i$ & plant state, input, and output \\ \bottomrule
\end{tabular}
\caption{Variable notations ($p$ denotes \emph{lo} or \emph{hi} precision).}
\label{tab:optimization_parameters}
\vspace{-4.mm}
\end{table}

\paragraph{\textbf{User-Defined:}}  
The variables \( A \), \( B \), \( C \), and \( K \) denote the discrete-time system and feedback gain matrices; \( h \) is the controller sampling period; and \( Q \) and \( R \) are the weighting matrices used in the LQR cost, \( T_{s} \) is the settling time and $\delta$ is the error bound that defines the reference band around the output reference.

The total simulation time \( T \), occurrence time instants of step inputs \( t_{j} \) (with a total of $r$ step inputs), and corresponding reference values \( \gamma_{j} \) are specified according to the controller design criteria. After each step input applied at \( t_{j} \), the plant output is expected to reach the reference \( \gamma_{j} \) within the specified settling time \( T_{s} \).  
The execution time of a control sample is denoted by \( t_{p} \), where \( p \in \{\emph{lo}, \emph{hi}\} \) indicates whether the iteration is executed with \emph{lo} or \emph{hi}-precision arithmetic, respectively.

\paragraph{\textbf{Precomputed Inputs:}} 
Switching between precisions is permitted only within specific time intervals represented by $[L_\beta, U_\beta] \in \mathcal{I}$, where each $[L_\beta, U_\beta]$ defines a \textit{range of samples} during which switching is allowed.
The construction of $\mathcal{I}$ and its cardinality $\mu = |\mathcal{I}|$ is based on the system's settling, peak, and rise times. In addition, $e_p$ denotes the sound roundoff error bounds associated with precision $p$, precomputed using state-of-the-art error analysis tools~\cite{fptaylor,daisy}.  
We describe the computation of these precomputed quantities and the construction of $\mathcal{I}$ in detail in~\autoref{subsec:precomputation}.

\paragraph{\textbf{Decision Variables:}} The decision variable $sw$ determines the precision type used at each sample.
For the $i$-th sample, $sw_i = 1$ indicates \emph{hi}-precision execution, and $sw_i = 0$ indicates \emph{lo}-precision execution. With $N = \lceil \frac{T}{h} \rceil$ samples over the simulation horizon $T$, our formulation includes a total of $N$ decision variables.

\paragraph{\textbf{Other Variables:}} 
The plant state ($x$), control input ($u$), and output ($y$) vectors are computed based on the current precision type determined by $sw$. At each sample $i$, these quantities are represented as $x_i$, $u_i$, and $y_i$.  
Depending on the value of $sw_i$, the corresponding system evaluations (see \autoref{eq:discrete-state-eqtn} and \ref{eq:control input}) are performed using \emph{hi}-precision for $sw_i = 1$ and \emph{lo}-precision for $sw_i = 0$.

%%%%%%%%%%%%%%%%%%%%%%% SECTION:  Overview of the Optimization Problem  %%%%%%%%%%%%%%%%%%%%%%%
\subsubsection{Problem Formulation}
\label{subsec:opt_overview}
\begin{figure}[t]
    \centering
    \begin{equation}\nonumber
        \boxed{
            \begin{array}{@{}l@{}l@{}l}
                \text{\textbf{minimize:}}  & z_{1}  +  z_{2}  \\
                 \text{where,} & z_{1}  =   \sum\limits_{i=0}^{N} \ ( \ sw_{i} \times t_{hi} + (1-sw_{i})\times t_{lo} \ )\\
                 & z_{2} = \sum\limits_{i=0}^{N} \ ( x_{i}^T \ Q \ x_{i} + u_{i}^T \ R \ u_{i} ) \\
                \text{\textbf{subject to:}}& & \\
                C: (\text{\textbf{settling-time}}) \ : 
                & \gamma_{j} - \delta  \ \le \ y_{i} \ \le \ \gamma_{j} + \delta,  \\          
                \text{where,} & \lceil \frac{ t_{j} + T_{s}}{h} \rceil \le i \le \lceil \frac{t_{j+1}}{h} \rceil, \ \  1 \le j \le r \\
            \end{array}
        }
    \end{equation}
    \caption{The MIQP formulation with non-linear constraints}
    \label{fig:opt_prob1}
\end{figure}
 
We present the complete optimization formulation in~\autoref{fig:opt_prob1}. 
The objective minimizes two quantities: (1) total execution time ($z_1$) and (2) the LQR cost ($z_2$) as defined in~\autoref{eq:lqr_cost}. 
The per-sample execution time depends on the selected precision---$t_{lo}$ for \emph{lo}-precision ($sw_i = 0$) and $t_{hi}$ for \emph{hi}-precision ($sw_i = 1$)---so $z_1$ sums $sw_i \times t_{hi} + (1 - sw_i) \times t_{lo}$ over all $N$ samples. 
Since both quantities are positive, the combined objective, $z_1 + z_2$, jointly balances execution time and control performance. 

The optimization is constrained by the settling-time condition ($C$): after each step input at time $t_j$, the plant output $y_i$ must reach and remain within its reference band $[\gamma_j - \delta, \gamma_j + \delta]$ by $t_j + T_s$, where $\gamma_j$ is the reference output and $T_s$ the settling time.

Finally, for each step input in $1 \le j \le r$, this constraint must hold for every discrete sample $i$ in the interval
\[
\left\lceil \frac{t_j + T_s}{h} \right\rceil \le i \le \left\lceil \frac{t_{j+1}}{h} \right\rceil,
\]
where $h$ is the controller's sampling period. 
This ensures that once the system has settled after a step input at $t_j$, the output $y_i$ remains within the reference band until the next step input occurs at $t_{j+1}$.
For the last step input at $t_r$, we define $t_{r+1} = T$, implying that the final reference $\gamma_r$ is maintained from $t_r + T_s$ until the end of the simulation horizon $T$.

%%%%%%%%%%%%%%%%%%%% SECTION: Formulation of Non-Linear Switching Constraints %%%%%%%%%%%%%%%%%%%%%%%%%%%%%%%
\subsubsection{Switching Constraints}
\label{subsec:opt_non_linear}
The main formulation in~\autoref{fig:opt_prob1} introduces the switching decision variable $sw_i$. 
We now formalize its behavior through additional switching constraints, presented in~\autoref{fig:opt_prob2}.

%%%%%% Optimization problem: Step 2 %%%%%%%%%%%%%%
\begin{figure}[t]
    \centering
    \begin{equation}\nonumber
        \boxed{
            \begin{array}{@{}l@{}l@{}l} 
                C_1: 
                & \sum\limits_{L_\beta \ \le \ i \ \le \ U_\beta}   ( sw_{i-1} \oplus sw_{i} ) \le 1, \\ 
                C_2: & \ sw_{i}\ = \ sw_{ \ U_\beta}, \ \ \ i \in [U_\beta +1, L_{\beta+1} - 1] \\ 
                C_3: 
                & \;\ sw_{i} = 0  \Rightarrow 
                (x_{i} = x_{i} +e_{lo}, \ \ u_{i} = u_i + e_{lo}, \ \ y_{i} = y_i + e_{lo}) \\
                C_4: 
                & \;\ sw_{i} = 1  \Rightarrow  
                 (x_{i} = x_i + e_{hi}, \ \ u_{i} = u_i + e_{hi}, \ \ y_{i} = y_i + e_{hi})
            \end{array}
        }
    \end{equation}
    \caption{Non-linear switching constraints}
    \label{fig:opt_prob2}
\end{figure}

\textbf{Constraint $C_1$} ensures that at most one precision switch occurs within each interval $[L_\beta, U_\beta] \in \mathcal{I}$, though it is also possible that no switching takes place. 
Formally, a switch in precision—either $(sw_{i-1}, sw_i) = (0,1)$ or $(1,0)$—may occur at most once for any sample $i \in [L_\beta, U_\beta]$. 
If a switch already occurs at the first sample of an interval (i.e., at $L_{\beta}$), then no other switching is allowed for the remaining samples in $[L_\beta + 1, U_\beta]$. 
Constraint $C_1$ captures this behavior, and it applies to all switching intervals in~$\mathcal{I}$.

Recall our running example of \texttt{CC} from \autoref{subsec:intro_example}. 
If the first switching interval is $[L_1, U_1] = [64, 120]$, constraint $C_1$ is given by:
$$
\sum_{64 \le i \le 120} (sw_{i-1} \oplus sw_i) \le 1.
$$
Here, the term $sw_{63} \oplus sw_{64}$ accounts for a switch at the interval start; if $(sw_{63}, sw_{64}) = (0,1)$ or $(1,0)$, then no further switch is permitted for $65 \le i \le 120$. 

\textbf{Constraint $C_2$} ensures that switching occurs only within the predefined intervals $[L_\beta, U_\beta]$ and once the interval $[L_\beta, U_\beta]$ ends, the switching variable remains unchanged until the next switching interval begins, i.e., it remains unchanged for all samples $i \in [U_{\beta}+1, L_{\beta+1}-1]$. This implies that the precision type remains the same during all those samples.

For the running example of \texttt{CC}, if the first switching interval is $[L_1, U_1] = [64, 120]$ (i.e., considering for $\beta=1$) and the next begins at $L_2 = 180$, the constraint $C_2$ is expressed as:
\[
sw_i = sw_{120}, \; \forall \, 121 \le i \le 179,
\]
and this holds for all switching intervals $[L_\beta, U_\beta]$ in $\mathcal{I}$.

Due to switching between precisions, each sample must account for the corresponding roundoff errors in the state, input, and output variables. 
Accordingly, \textbf{constraints $C_3$ and $C_4$} incorporate the appropriate error terms ($e_{lo}$ or $e_{hi}$) into the state-update and control equations, depending on the precision selected. 
These equations apply to all samples $1 \le i \le N$.

%%%%%%%%%%%%%%%%%%%%%%%%%%%% SECTION: Linearization of Constraints %%%%%%%%%%%%%%%%%%%%%%%%%%%%%%%
\subsubsection{Linearization of Constraints}
\label{subsec:opt_linear}
The switching constraints in \autoref{fig:opt_prob2} introduce non-linearity into the optimization problem in two ways: first, through the `XOR' ($\oplus$) operation in constraint $C_1$, and second, through the `implication relations' in constraints $C_3$ and $C_4$. 
To ensure tractability, we linearize these constraints exactly, without introducing any over-approximations. 
\autoref{fig:opt_prob3} presents our linearized formulations for the optimization problem.
 
%%%%%% Optimization problem: Step 3 %%%%%%%%%%%%%%
\begin{table*}[h]
    \small
    \centering
    \begin{tabular}{c|c|c|c|c|c|c|c}
       \toprule
       \boldmath$sw_{i-1}$  &  \boldmath$sw_{i}$ & \boldmath$sw_{i-1} \oplus sw_{i}$ & \boldmath$sw_{i-1} - sw_{i}$ & \boldmath$sw_{i} - sw_{i-1}$ & \boldmath$2 - (sw_{i-1} + sw_{i})$ & \boldmath$(sw_{i-1} + sw_{i})$  & \boldmath$\chi_i$  \\ \midrule
        $0$ & $1$ & $1$ & $-1$ & $1$ & $1$ & $1$ & $1$ \\ 
        $1$ & $0$ & $1$ & $1$ & $-1$ & $1$ & $1$ & $1$ \\ 
        $0$ & $0$ & $0$ & $0$ & $0$ & $2$ & $0$ & $0$ \\ 
        $1$ & $1$ & $0$ & $0$ & $0$ & $0$ & $2$ & $0$ \\ \bottomrule
    \end{tabular}
    \caption{Linearization of `XOR'}
    \label{tab:XOR-linearize}
\end{table*}
\begin{figure}[t]
    \centering
    \begin{equation}\nonumber
        \boxed{
            \begin{array}{@{}l@{}l@{}l}
                C_{1}(1) \ : 
                &  \chi_{i} \ge sw_{i-1} - sw_{i}, \\
                C_{1}(2) \ : 
                &  \chi_{i} \ge sw_{i} - sw_{i-1},\\ 
                C_{1}(3) \ : 
                &  \chi_{i} \le 2 - (sw_{i-1} + sw_{i}), \\
                C_{1}(4) \ : 
                &  \chi_{i} \le (sw_{i-1} + sw_{i})   \\  
                C_{1}(5) \ : 
                & \sum\limits_{L_\beta \ \le \ i \ \le \  U_\beta} \  \chi_{i}  \le 1 \\ \\

                C_{3,4}(1) \ : 
                & \ e_i = e_{lo} \times (1-sw_i) +  e_{hi} \times sw_i \\
                C_{3,4}(2) \ : 
                & \ x_i = x_i + e_i  \\
                C_{3,4}(3) \ : 
                & \ u_i = u_i + e_i  \\
                C_{3,4}(4) \ : 
                & \ y_i = y_i + e_i  \\
            \end{array}
        }
    \end{equation}
    \caption{MIQP with linear constraints}
    \label{fig:opt_prob3}
\end{figure}

Constraints $C_{1}(1)$–$C_{1}(4)$ linearize the XOR operation from constraint $C_1$ by introducing a new binary variable $\chi_i$. 
Table~\ref{tab:XOR-linearize} illustrates this linearization.

Column~3 lists the expected XOR values, while Columns~4–7 show the corresponding evaluations for 
$sw_{i-1} - sw_i$, $sw_i - sw_{i-1}$, $2 - (sw_{i-1} + sw_i)$, and $(sw_{i-1} + sw_i)$, respectively. 
Column~8 reports the resulting $\chi_i$ values obtained using constraints $C_{1}(1)$–$C_{1}(4)$, e.g., when $sw_{i-1} = 0$ and $sw_i = 1$, the constraints imply $\chi_i \ge -1$, $\chi_i \ge 1$, $\chi_i \le 1$, and $\chi_i \le 1$, which together yield $\chi_i = 1$. 
Repeating this for all cases verifies that $\chi_i = sw_{i-1} \oplus sw_i$ holds for every $i$. 
Finally, $C_{1}(1)-C_{1}(5)$ replace the original $C_1$ in the optimization problem.

Constraints $C_{3,4}(1)$–$C_{3,4}(4)$ linearize the implication relations in $C_3$ and $C_4$. 
Constraint $C_{3,4}(1)$ defines $e_i$ as the precision-dependent error term that holds $e_{hi}$ for \emph{hi}-precision ($sw_i = 1$) or $e_{lo}$ for \emph{lo}-precision ($sw_i = 0$). 
The remaining constraints incorporate $e_i$ into the state, control, and output equations to capture the roundoff effects consistently across all samples.

This completes the linearization of all non-linear constraints, resulting in an MIQP formulation with entirely linear constraints.

%%%%%%%%%%%%%%%%%%%%%%%%%%%%%%%%  SECTION:   Precision Switching %%%%%%%%%%%%%%%%%%%%%%%%%%%%

\subsection{Precomputed Inputs}
\label{subsec:precomputation}
In the previous sections, we formulated the optimization problem assuming that the switching intervals and roundoff error bounds were available as precomputed inputs. This section details the pre-computation phase, where these quantities are computed before being supplied to the optimization phase.

\subsubsection{Designing Switching Intervals}
\label{subsec:switching}
We begin by describing the pre-design of the switching intervals, collectively represented as $\mathcal{I} = \{ [L_{\beta}, U_{\beta}] \}_{\beta=1}^{\mu}$ 
(see~\autoref{tab:optimization_parameters}). 
A switch between \emph{hi} and \emph{lo} precisions is allowed at most once within each interval $[L_{\beta}, U_{\beta}]$, where $1 \le \beta \le \mu$, $\mu$ being the total number of intervals. 
These intervals are designed based on the system's time-domain performance metrics, \emph{rise time} ($T_r$), \emph{peak time} ($T_p$), and \emph{settling time} ($T_s$), to capture regions where precision has the most impact on control performance.

\paragraph{Motivation}
Consider a control system with a settling time $T_s = \SI{1}{\second}$, output reference $\gamma = 0.5$, and an error margin of $5\%$. 
The output is expected to reach and remain within $[0.45, 0.55]$ within $\SI{1}{\second}$. 
However, when implemented using \emph{lo}-precision arithmetic, the output may only reach $0.432$ at $\SI{1}{\second}$, failing to meet the settling-time requirement due to large roundoff errors. 
Such inaccuracies often increase the effective settling time, potentially  leading to prolonged deviations from the desired setpoint. 

Since shorter settling, peak, and rise times ($T_s$, $T_p$, $T_r$) indicate fast response and better performance (see~\autoref{sec:background}), it is desirable to adapt the controller precision accordingly. 
Hence, we construct the switching intervals based on these metrics.

\paragraph{Intuition}
Immediately after a step input, the system exhibits its largest deviation from the desired reference, both during the initial rise (until approximately $50\%$ of $T_r$) and near the peak overshoot before settling. 
During these periods, precise control computations are crucial for maintaining performance requirements and achieving fast convergence. 
Hence, a switch from \emph{lo} to \emph{hi} precision is most beneficial within these critical windows. 
This intuition guides the construction of switching intervals in $\mathcal{I}$.
%Suppose, the outputs are $0.432$ and $0.476$ at $\SI{1}{\second}$, when LO and HI controllers are used, respectively. This indicates that the output fails to reach the desired reference within the settling time, when the LO controller is used. 

%A shorter settling time is desired since it indicates that the system quickly stabilizes towards its reference, after facing a perturbation at its input. Similarly, shorter peak and rise time make the system more responsive towards step inputs. These can be guaranteed with more accurate control outputs, generally obtained with higher-precision arithmetic. 

\paragraph{Interval Construction}
For each step input applied at $t_j$ ($2 \le j \le r$), we construct two switching intervals:
\begin{compactenum}
\item One interval that covers the period from the step input instant to half of the rise time: \[
    [L_{2j-2}, U_{2j-2}] = 
    \left[ 
    \left\lceil \frac{t_j}{h} \right\rceil, 
    \left\lceil \frac{t_j + T_r/2}{h} \right\rceil 
    \right].
    \]
\item The other interval covers the time from the first peak overshoot to when the system settles:
    \[
    [L_{2j-1}, U_{2j-1}] = 
    \left[
    \left\lceil \frac{t_j + T_p}{h} \right\rceil,
    \left\lceil \frac{t_j + T_s}{h} \right\rceil
    \right].
    \]
\end{compactenum}
The system starts with a step input, in a perturbed state at $t_1 = 0$. 
The first interval corresponds to the initial transient response and is defined as:
\[
[L_1, U_1] = \left[\left\lceil \frac{T_p}{h} \right\rceil, \left\lceil \frac{T_s}{h} \right\rceil \right],
\]
indicating that the initial samples (till $T_p$) are executed using \emph{hi}-precision arithmetic (as the system is far from steady state) and a switch is first allowed at the sample $\left\lceil T_p/h \right\rceil$.
Overall, for $r$ step inputs, we obtain $2r - 1$ switching intervals in total.

We now illustrate the interval construction for our running example of \texttt{CC} 
from ~\autoref{subsec:intro_example}.
The system's time-domain performance metrics are: settling time $T_s = \SI{1.2}{\second}$, peak time $T_p = \SI{0.64}{\second}$, and $50\%$ of the rise time, $T_r/2 = \SI{0.3}{\second}$. 
The sampling period is $h = \SI{10}{\milli\second}$. 

\begin{figure}[t]
    \centering
    \includegraphics[scale=0.43]{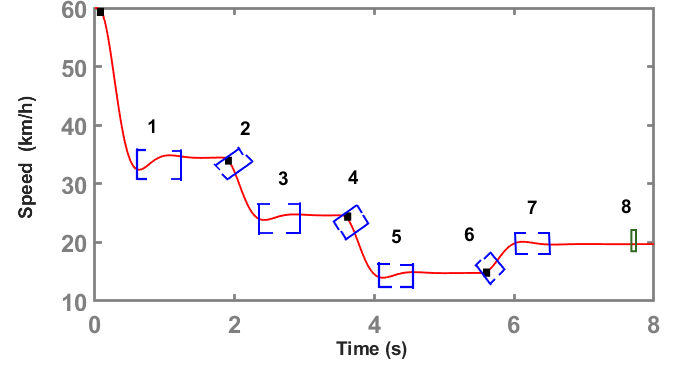}
    \caption{Switching intervals for the cruise control (\texttt{CC}) system}
    \label{fig:CC_switch}
\end{figure}

Let us assume that four step inputs are applied at $\SI{0}{\second}$, $\SI{1.8}{\second}$, $\SI{3.5}{\second}$, and $\SI{5.5}{\second}$.
Using the construction method described above, the switching intervals are:
\[
\begin{aligned}
1) [L_{1}, U_{1}] &= [64, 120] & (\text{for } t_1 = 0),\\
2) [L_{2}, U_{2}] &= [180, 210], \quad 3) [L_{3}, U_{3}] = [244, 300] & (\text{for } t_2 = 1.8),\\
4) [L_{4}, U_{4}] &= [350, 380], \quad 5) [L_{5}, U_{5}] = [414, 470] & (\text{for } t_3 = 3.5),\\
6) [L_{6}, U_{6}] &= [550, 580], \quad 7) [L_{7}, U_{7}] = [614, 670] & (\text{for } t_4 = 5.5).
\end{aligned}
\]
These seven switching intervals are shown in ~\autoref{fig:CC_switch} (highlighted in blue).

\paragraph{Additional Switching}
While the intervals above cover precision switching during perturbed phases, it can also be beneficial to allow limited switching once the system gets settled. 

In the example, the last step input occurs at $t_r=\SI{5.5}{\second}$, and the system reaches its reference by approximately $\SI{6.7}{\second}$ (i.e., $t_r + T_s$). 
The simulation continues until $\SI{8}{\second}$, when the output remains stable, within the reference band. 
Executing only \emph{hi}-precision samples for the remaining time would be wasteful, while running entirely in \emph{lo}-precision could degrade performance. 
Thus, one or more switching(s) are permitted after the system settles to balance runtime and control performance.

To model this, we allow more switching opportunities every $T_s$ seconds after the system has remained settled for an additional duration of $T_s$, i.e., starting from sample $
\tau = \left\lceil \frac{t_r + 2T_s}{h} \right\rceil
$.
Specifically, a switching interval is designed every $\lceil T_s/h \rceil$ samples thereafter:
\[
[\tau + k \lceil T_s/h \rceil, \, \tau + k \lceil T_s/h \rceil + 1], \quad k = 0, 1, 2, \dots
\]
until the total simulation time $T$ is reached.
Hence, the total number of switching intervals is: $\mu = (2r - 1) + \left\lceil \frac{T - (t_r + 2T_s)}{T_s} \right\rceil$.

For the CC system, $\mu = 8$, where the additional switching interval is $[L_8, U_8] = [790, 791]$, shown in green in ~\autoref{fig:CC_switch}. 
This final interval enables adaptive precision use even after the transient response has subsided, maintaining high efficiency without sacrificing control performance.

\subsubsection{Computing Roundoff Errors}
To account for roundoff errors in each controller execution, we precompute them using FPTaylor~\cite{fptaylor}, a state-of-the-art tool for sound worst-case roundoff error analysis of floating-point programs.
FPTaylor applies global optimization techniques to soundly maximize roundoff error expressions based on the given bounds of input variables, thereby providing mathematically rigorous upper bounds on the roundoff errors for a specified floating-point precision.

In our setup, FPTaylor requires bounds on the values of the plant state variables ($x$) and control inputs ($u$) to compute these errors.
Since these quantities represent physical parameters of the control system, their feasible ranges can typically be estimated from the system design, simulation results, or prior domain knowledge.
We assume that such ranges are known or can be conservatively estimated—a reasonable assumption in practice, as control engineers generally have access to these limits during controller design.
Using these ranges, we compute sound roundoff error bounds for both \emph{hi}- and \emph{lo}-precision floating-point implementations. However, we assume the inputs are represented exactly; therefore, we do not consider any roundoff error in the input.

For the running example of \texttt{CC}, the system has three state variables and one control input. We obtained the bounds on these variables directly from the control model, design specifications, and through simulations. For 16-bit and 32-bit floating-point precision levels, the roundoff error bounds were computed as $1.46 \times 10^{-1}$ and $1.74 \times 10^{-5}$, respectively. These values are then used as precomputed inputs to the optimization problem.

%%%%%%%%%%%%%%%%% IMPLEMENTATION %%%%%%%%%%%%%%%%%%
\subsection{Implementation}
\label{subsec:implementation}
We have implemented our approach in a prototype tool, which will be released as open source upon acceptance.
The tool takes as input user-defined parameters (see~\autoref{tab:optimization_parameters}) and system time-domain performance metrics (rise, peak, and settling times) and outputs the \emph{optimal precision-switching schedule} for the given controller.
It automatically constructs the switching intervals from the timing properties, generates the MIQP formulation, invokes the solver, and computes the optimal schedule.
Roundoff errors are computed by providing bounds on the state variables and control inputs of the system and invoking FPTaylor.
Currently, this step is carried out manually, but full integration into the tool can be straightforwardly implemented.

\paragraph{Choice of Tools}
We integrated our tool with Gurobi~\cite{gurobi} as the \emph{MIQP solver} to compute the optimal precision-switching schedule. Gurobi is a high-performance industrial solver known for its speed, robustness, and scalability, and it efficiently handles convex MIQP problems where the objective function is quadratic and the constraints are linearized, which aligns with our formulation.

Gurobi internally uses 64-bit floating-point precision, with numerical tolerances typically at $10^{-9}$.
While our approach is conceptually capable of supporting arbitrary precision levels, this inherent tolerance bound limits the current tool to at most 32-bit floating-point controller implementations.
At higher precisions, such as 64-bit, the magnitude of roundoff errors becomes comparable to or smaller than the solver's tolerance, leading to unreliable optimization behavior or convergence issues.

Other industrial solvers (e.g., CPLEX~\cite{cplex_code}) operate under similar tolerances and therefore face the same limitation.
The SCIP Optimization Suite~\cite{scip} supports extended-precision arithmetic, achieving tolerances up to $10^{-15}$, which offers slightly improved numerical stability but at a significant cost in computational efficiency. Even then, such tolerances may still be insufficient to reliably handle 64-bit implementations. Nevertheless, 32-bit floating-point precision is widely used for embedded and control applications, making this design choice both realistic and practically sufficient.

For \emph{roundoff error computation}, we use FPTaylor~\cite{fptaylor}, a state-of-the-art global optimization–based tool that provides sound worst-case error bounds, often more precise than other existing techniques.
In principle, however, any tool capable of computing sound worst-case roundoff error bounds can be employed.

%% file: sections/evaluation.tex
\section{Evaluation}
\label{sec:evaluation}
In this section, we evaluate our approach based on the following three research questions:
\begin{compactenum}[RQ1.]
    \item 
    How does control performance and runtime vary across different precision baselines (using \emph{hi}- and \emph{lo}-precision overall)?
    \item For a given control system, what proportion of \emph{lo}-precision samples achieves the best trade-off? 
    \item How efficient is the optimized switching schedule compared to the baselines in terms of runtime and control performance?  
\end{compactenum}

%%%%%%%%%%%%%%%%%%%%%%%%%%  Table for details of the benchmarks %%%%%%%%%%%%%%%%%%%%%%

\paragraph{Benchmarks} 
We evaluate our method using 9 benchmark control systems from the automotive domain, with system orders (i.e., number of state variables) ranging from 2 to 5.
Each benchmark models a feedback control system, which we briefly explain below.

We consider 4\textbf{ second-order} systems:
1) the \emph{DC-motor speed controller} (\texttt{MS})~\cite{roy2016multi}, which regulates motor speed by adjusting terminal voltage;
2) the \emph{F1-Tenth lane-following controller} (\texttt{F1})~\cite{banerjee_P2SDS}, which keeps a small autonomous car aligned with the lane center by controlling its steering angle;
3) the \emph{resistor–capacitor circuit} (\texttt{RC})~\cite{banerjee_P2SDS}, which models voltage regulation for signal filtering; and
4) the \emph{DC-servo controller} (\texttt{DCS})~\cite{online2018stability}, which adjusts servo acceleration to regulate angular position;

The 2 \textbf{third-order} systems are:
1) the \emph{cruise controller} (\texttt{CC})~\cite{roy2016multi}, which maintains vehicle speed by modulating throttle force; and
2) the \emph{adaptive cruise controller} (\texttt{ACC})~\cite{online2018stability}, which regulates spacing error with respect to a leading vehicle.

The 2 \textbf{fourth-order} systems include:
1) the \emph{lane-keeping controller} (\texttt{LK})~\cite{online2018stability}, which minimizes lateral position error by computing appropriate steering inputs; and
2) the \emph{suspension controller} (\texttt{SC})~\cite{roy2016multi}, which stabilizes the vehicle’s vertical position by adjusting input forces.

Finally, a \textbf{fifth-order} system is considered, which is the \emph{vision-based lateral controller} (\texttt{LC})~\cite{LC_system}, which uses image-based feedback to compute the front-wheel steering angle required to maintain the vehicle’s position within the lane.

The \texttt{ACC} and \texttt{CC} are \emph{open-loop unstable} systems, which become completely unstable without the control input from their feedback controllers~\cite{ogata2020controlBook}. In contrast, the remaining are \emph{open-loop marginally stable}, operating around an unstable equilibrium where without the control input they neither converge to the reference nor diverge but exhibit sustained oscillations~\cite{ogata2020controlBook}. In both cases, the feedback control in the closed loop stabilizes the system.

\paragraph{Experimental Setup}
We have considered three execution modes of the controller: 1) full 16-bit floating-point execution (FP16) as \emph{lo}-precision,
2) full 32-bit floating-point execution (FP32) as \emph{hi}-precision, and
3) generated  precision-switching schedule, where the controller switches between FP16 and FP32. 

All experiments for generating optimal schedules were conducted on a 64-bit Windows system with an Intel Core i3 processor (1.20 GHz) and 8 GB of RAM.
We used Gurobi 12.0.2 as the external MIQP solver and FPTaylor v0.9.3 for sound roundoff error analysis.
For simulating the generated schedules, we used an NVIDIA GPU with 24 GB of RAM and Tensor Core support, which provides native hardware support for both FP16 and FP32 arithmetic.

\subsection{RQ1: FP16 vs FP32}
\label{subsec:exp1_baselines}
We begin by comparing the control performance in terms of the \emph{LQR cost}, for two baselines: controller fully executed in (1)  FP16 and (2) FP32.
Table~\ref{tab:baselines_output} summarizes the results. Column 1 and 2 present all the  benchmarks with their sampling periods ($h$) and settling times ($T_s$), and the minimum and maximum values of output references, respectively.
Column~5 presents the percentage reduction in LQR cost, where positive values indicate an improvement with FP32 over FP16 execution.

%%%%%%%%%%%%%%% Table for baselines %%%%%%%%%%%%%%%%%
\begin{table}[t]
\small
\renewcommand{\arraystretch}{1.2}
    \centering
    \begin{tabular}{c|c|cc|c}
        \toprule       
         \textbf{benchmark} & \textbf{ref. output} & \multicolumn{3}{c}{\textbf{LQR cost}} 
        \\ \cline{3-5}        
        \boldmath$(h, T_s)$ & \textbf{(min, max)} & \multicolumn{1}{c|}{\textbf{FP16}}  & \textbf{FP32}  &  {\textbf{reduction\%}}\\ \midrule
        %% MS
         \texttt{MS} \ ($\SI{20}{\milli\second}$, $\SI{0.6}{\second}$) & $(0.00,1.50)$ & \multicolumn{1}{c|}{$856.0$} & $572.3$ & $33.1$  \\ 
        %% F1
        \texttt{F1} \ ($\SI{20}{\milli\second}$, $\SI{0.6}{\second}$) & $(0.10,0.40)$ & \multicolumn{1}{c|}{$32.0$} & $31.0$ &  $3.2$  \\
        %% RC
        \texttt{RC} \ ($\SI{20}{\milli\second}$, $\SI{0.3}{\second}$) & $(0.00,5.00)$ & \multicolumn{1}{c|}{$669.5$} & $419.7$ & $37.3$  \\
        %% DCS
        \texttt{DCS} \ ($\SI{10}{\milli\second}$, $\SI{0.8}{\second}$) & $(4.00,10.00)$ & \multicolumn{1}{c|}{\textcolor{red}{\textit{Inf}}} & $2561960.0$ & ---  \\
        %% CC
        \texttt{CC} \ ($\SI{10}{\milli\second}$, $\SI{1.2}{\second}$) & $(15.00,60.00)$ & \multicolumn{1}{c|}{\textcolor{red}{\textit{NaN}}} & $341106.0$ & ---  \\
         %% ACC
         \texttt{ACC} \ ($\SI{20}{\milli\second}$, $\SI{0.8}{\second}$) & $(1.00,7.00)$ & \multicolumn{1}{c|}{\textcolor{red}{\textit{Inf}}} & $575550.0$ & ---  \\
        %% LK
        \texttt{LK} \ ($\SI{20}{\milli\second}$, $\SI{0.8}{\second}$) & $(0.03,1.00)$ & \multicolumn{1}{c|}{$48.1$} & $42.5$ & $11.6$  \\
        %% SC
        \texttt{SC} \ ($\SI{20}{\milli\second}$, $\SI{0.4}{\second}$) & $(0.03,0.50)$ & \multicolumn{1}{c|}{$58.5$} & $56.5$ & $3.6$  \\
        %% LC
        \texttt{LC} \ ($\SI{10}{\milli\second}$, $\SI{0.8}{\second}$) & $({\text -}0.08,0.08)$ & \multicolumn{1}{c|}{$29.3$} & $28.1$ & $4.1$  \\ \bottomrule
    \end{tabular}
    \caption{FP16 vs FP32 in terms of LQR costs}
    \label{tab:baselines_output}
\end{table}

\begin{figure}[t]
    \centering
    \includegraphics[scale=0.35]{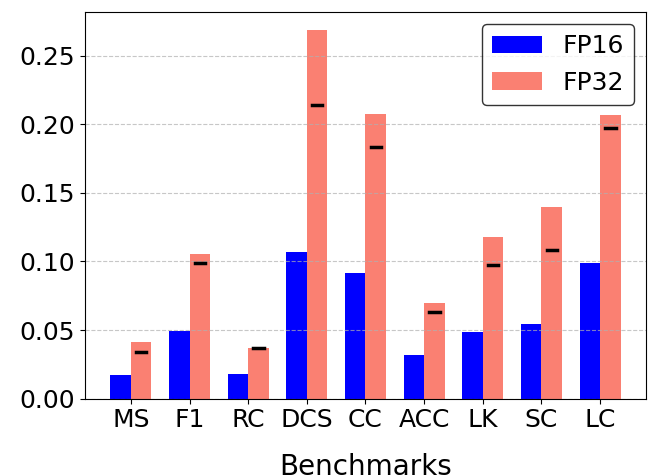}
    \caption{FP16 vs FP32 in terms of runtime}
    \label{fig:FP16_vs_FP32_runtime}
\end{figure}

Our results show that across \emph{all} benchmarks, FP16 execution leads to significantly high LQR costs, indicating degraded control performance. In particular, the open-loop unstable systems (\texttt{ACC} and \texttt{CC})
show catastrophic control performances under FP16, producing \emph{Inf} and \emph{NaN} LQR costs due to overflows and severe roundoff error accumulation.

We observe that these systems have high LQR costs even under FP32, as their dynamics demand larger control efforts to reach and maintain the desired reference. 
Moreover, \texttt{ACC} has higher cost than \texttt{CC} primarily because \texttt{CC} uses a shorter sampling period (see Column 1 of \autoref{tab:baselines_output}), enabling more frequent control execution and a faster closed-loop response.

For the remaining open-loop marginally stable systems, the control effort required to reach the reference is smaller, resulting in substantially lower LQR costs (except for \texttt{DCS}). The magnitude of the system output variable and other state variables also affect the scale of the LQR cost: e.g., for \texttt{F1} system, the (min, max) values of the reference output are $(0.1, 0.4)$ and the LQR cost obtained is $31.0$, whereas the cost is $419.7$ for \texttt{RC} with the (min, max) as $(0.0, 5.0)$.

The \texttt{DCS} system, however, is an exception. Despite being marginally stable with a short sampling period and settling time, it exhibits degraded control performance in FP16 resulting in \emph{Inf} LQR cost. However, it follows the settling time requirement in FP32, having a large but finite LQR cost. 
This behavior likely arises from its inherent system dynamics, which require strong control effort, thereby leading to high LQR costs.

Overall, FP32 consistently yields finite and substantially lower costs across all benchmarks, regardless of whether the system is open-loop unstable or marginally stable, with the largest improvement of 37.3\% for \texttt{RC} and the smallest of 3.2\% for \texttt{F1}, demonstrating stable control performance overall.

Next, we compare FP16 and FP32 in terms of runtime. For each benchmark, we measured the average runtime for all samples over 3 runs for both FP16 and FP32 executions. ~\autoref{fig:FP16_vs_FP32_runtime} presents the results. As expected, for \emph{all} benchmarks, FP16 execution is consistently much faster, while FP32 takes twice as long to complete the same control execution for all benchmarks (the black markers in the FP32 bar charts indicate twice the FP16 runtime).

In summary, these experiments show that FP32 consistently delivers superior control performance compared to FP16 across all benchmarks, albeit at a significantly higher computational cost. This indicates substantial potential for achieving a balanced trade-off between performance and runtime efficiency — a goal we aim to achieve through our proposed precision-switching schedule that alternates between FP16 and FP32. However, for benchmarks where even FP32 yields large LQR costs (such as \texttt{DCS} and \texttt{ACC}), introducing FP16 execution, even for a small number of samples, may lead to  degraded control performance due to accumulated roundoff errors. For other systems, however, an optimized precision-switching schedule can effectively balance both control quality and runtime, achieving the best of both precision options.

%% file: sections/rq2.tex
\subsection{RQ2: Optimal Precision-Switching Schedule}
\label{subsec:exp2_optimization_output}
We now evaluate the effectiveness of our MIQP-based precision-switching scheduling approach through the optimal switching schedules it generates for all benchmarks.

Each system's parameters, such as sampling period, settling time, output reference band, and total simulation time are configured according to its control design, that vary across all benchmarks. The total number of samples is derived from the sampling period and total simulation time. Wherever available, we use physical bounds on the state and input variables extracted directly from the control model or design specifications. For the remaining cases, we obtain feasible input bounds through simulation, which are then used to compute sound worst-case roundoff error bounds of the controller execution.

While the simulated bounds are not formally sound, they are conditioned on realistic operational ranges and provide sufficiently tight estimates for our evaluation. Since these values serve as inputs to our approach, having sound bounds on the variables would directly yield sound roundoff error bounds as well. This assumption is also practical, as control engineers typically know or can conservatively approximate such variable ranges during the controller design phase.

We measured the average per-sample runtime for each precision by executing the controller on a GPU over 1000 samples. These precomputed quantities, together with the system dynamics, form the inputs to our optimization problem. (We will release all experimental data together with the tool.)

\begin{table}[t]
\small 
\renewcommand{\arraystretch}{1.2}
    \centering
    \begin{tabular}{c|ccc|cc|c}
        \toprule
        \multirow{2}{*}{} 
        & \multicolumn{3}{c|}{\textbf{precomputed inputs}}
        & \multicolumn{2}{c|}{\textbf{outputs}} &  
        \\  \cline{2-4} \cline{5-6} 

        & \multicolumn{2}{c}{\textbf{roundoff (e)}} & \multirow{2}{*}{\boldmath$\mu$} & {\multirow{2}{*}{\boldmath$\#$\textbf{sw}}} & {\multirow{2}{*}{\textbf{\% FP16}}} & {\textbf{time (s)}} \\ 
        
        & \multicolumn{1}{c}{\textbf{FP16}} & {\textbf{FP32}} & &  & \textbf{samples} &  \\ \midrule

        \texttt{MS} & {2.27e-3} &  2.51e-7 & $4$ &  2 & $33.33$ & $0.22$ \\

         \texttt{F1} & {7.58e-4} &  7.86e-8 & $9$ & 5 & $85.75$ & $1.03$ \\
         
         \texttt{RC} & {9.13e-3} &  {1.08e-6} & $9$ &  2 & $86.00$ & $0.33$\\
        
        \texttt{DCS} & {7.25e-2} &  {9.11e-6}  & $10$ &  0 & $0.00$ & $1.72$ \\
         
        \texttt{CC} & {1.46e-1} &  {1.74e-5}  & $8$  &  6 & $27.75$ & $3.21$ \\
        
         \texttt{ACC} &  {1.72e-1} &  {1.87e-5}  &  $7$ &  0 & $0.00$ & $0.73$ \\
         
        \texttt{LK} &  {2.73e-2} &  {3.45e-6}  & $9$ &  6 & $26.25$ & $2.03$ \\
        
         \texttt{SC} &  {1.62e-1} &  {1.85e-5}  &  $13$ &  6 & $10.50$ & $2.35$ \\
         
        \texttt{LC} &  {1.94e-4} &  {5.20e-8}  & $7$ &  6 & $30.00$ & $6.84$ \\ \bottomrule 
        
    \end{tabular}
    \caption{Evaluation of our approach: precomputed inputs, generated switching schedules, and runtimes (averaged over 3 runs)}
    \label{tab:optimization_output}
    \vspace{-4mm}
\end{table}

~\autoref{tab:optimization_output} summarizes the results.
For each benchmark, Columns 2–4 list the precomputed inputs: Columns 2 and 3 report the roundoff errors in FP16 and FP32 obtained using FPTaylor, while Column 4 gives the total number of switching intervals ($\mu$) determined from the system’s control performance metrics.
Columns 5 and 6 present the outputs of the optimization problem: Column 5 shows the number of optimized precision switches (\#sw), and Column 6 reports the percentage of FP16 samples in the resulting schedule.
The last column provides the total runtime of the optimization process in seconds, including solver time, averaged over 3 runs.

%%%%%%%%%%%%%%%% Output of MIQP %%%%%%%%%%%%%%%%%
As expected, systems such as \texttt{DCS} and \texttt{ACC} produce schedules with no FP16 samples. This behavior aligns with the high LQR costs observed for FP32 and the \emph{Inf} results under FP16 in~\autoref{tab:baselines_output}, confirming that even executing a few low-precision samples can lead to degraded control performance. Interestingly, \texttt{CC} yields a valid switching schedule despite exhibiting \emph{NaN} in FP16. 
For \texttt{CC}, the relatively large FP16 roundoff errors (see~\autoref{tab:optimization_output}, Column~2) accumulate and produce invalid intermediate values during the LQR cost computation, leading to \emph{NaN} for all-FP16 execution. In contrast, the much smaller FP32 errors keep all computations well-defined, allowing the optimizer to safely use FP16 for a small portion of the samples (27.75\%).

Overall, our approach generates effective switching schedules for most benchmarks, with both precisions actively utilized. In particular, 6 out of 9 benchmarks exhibit reasonably balanced schedules, switching in 50\% or more of the available intervals ($\mu$). The \texttt{RC} system presents an interesting exception: despite having 9 switching intervals, it switches only 2 times, executing about 86\% of its samples in FP16. This behavior results from its short settling time and limited step inputs, allowing continuous low-precision execution. In contrast, for \texttt{SC}, the relatively high FP16 roundoff errors restrict FP16 usage to just 10.5\%.

These results show that our optimization effectively maximizes FP16 usage, thus identifying where low precision can be safely used, without compromising the control performance requirements.

We also find that our approach is computationally efficient, as shown in Column~7 of~\autoref{tab:optimization_output}. The reported times include both the formulation of the optimization problem and its solution using the Gurobi MIQP optimizer. For nearly all systems, the complete optimization takes under 5 seconds, with \texttt{LC} requiring the longest time of 6.84 seconds. The runtime scales with the system dynamics, number of samples and the switching intervals allowed. These results demonstrate that solving the MIQP problem with Gurobi is not a bottleneck, making our method a practical option for computing an optimal precision-switching schedule.

%% file: sections/rq3.tex
\subsection{RQ3: Our Approach vs Baselines}
\label{subsec:exp3_comparison_baselines}

We now compare the performance of our generated precision-aware switching schedules against two baseline configurations: (1) all-FP16 and (2) all-FP32 executions.
Since we did not find optimized switching schedules for \texttt{DCS} and \texttt{ACC}, these systems are excluded from this comparison.
Table~\ref{tab:baselines_comparison} shows the results. The LQR costs obtained using our schedules for all benchmarks are presented (Column~2), along with the percentage reduction relative to FP16 (Column~3) and relative to FP32 (Column~4).

Our results show that for \emph{all} benchmarks, the LQR costs for the switching schedule are substantially lower than that of the all-FP16 (Column~3). The highest reduction is 37.314\% for \texttt{RC}, while the smallest is 3.244\% for \texttt{F1}. For \texttt{CC}, the switching schedule  successfully produces a finite LQR cost, where the FP16 resulted in a \emph{NaN}.

\begin{table}[t]
    \small
    \renewcommand{\arraystretch}{1.2}
    \centering
    \begin{tabular}{c|ccc}
        \toprule

        \multirow{3}{*}{\textbf{benchmark}} & \multicolumn{3}{c}{\textbf{LQR cost}}  \\ \cline{2-4}

        & \multicolumn{1}{c|}{\textbf{switching}} 
        & \multicolumn{2}{c}{\textbf{\% reduction w.r.t.}}   \\

         & \multicolumn{1}{c|}{\textbf{schedule}} & \multicolumn{1}{c}{\textbf{FP16}} & \textbf{FP32} \\ \midrule

        %% MS
        \texttt{MS}  & \multicolumn{1}{c|}{$572.334$} & {$33.139$}  & $-0.0059$ \\

        %% F1
        \texttt{F1}  & \multicolumn{1}{c|}{$30.962$} & {$3.244$} & $0.1226$ \\

        %% RC
        \texttt{RC}  & \multicolumn{1}{c|}{$419.680$} & {$37.314$}  & $0.0048$ \\

        %% CC
        \texttt{CC}  & \multicolumn{1}{c|}{$341041$} & {--- } & $0.0190$ \\

        %% LK
        \texttt{LK}  & \multicolumn{1}{c|}{$42.541$} & {$11.546$}  & $-0.0965$ \\

        %% SC
         \texttt{SC}  & \multicolumn{1}{c|}{$56.452$} & {$3.552$}  & $0.0850$ \\ 

         %% LC
        \texttt{LC}  & \multicolumn{1}{c|}{$28.126$} & {$4.049$}  & $-0.0925$ \\ \bottomrule

    \end{tabular}
    \caption{LQR costs for the generated switching schedules and their reduction \% w.r.t. the baselines (-ve values indicate cost increases)}
    \label{tab:baselines_comparison}
    \vspace{-4mm}
\end{table}

Moreover, the LQR cost achieved by the switching schedule closely matches that of FP32, with cost increases typically below 0.10\% (Column~4) across \emph{all} benchmarks. This enhanced control performance results from our switching interval design which strategically enables precision transitions; allowing the optimized schedule to assign FP32 samples during transient phases (e.g., immediately after step inputs or near peak responses) and FP16 samples once the system reaches steady state. This scheduling effectively limits roundoff accumulation, with control performance on par with FP32.

Interestingly, for some systems (\texttt{F1}, \texttt{CC}, etc.), the LQR cost under the switching schedule is marginally lower than that of FP32. This effect can occur when rounding errors partially cancel each other, occasionally leading to results that are even more accurate than consistently using higher precision.

\begin{figure}[h]
    \centering
    \includegraphics[scale=0.3]{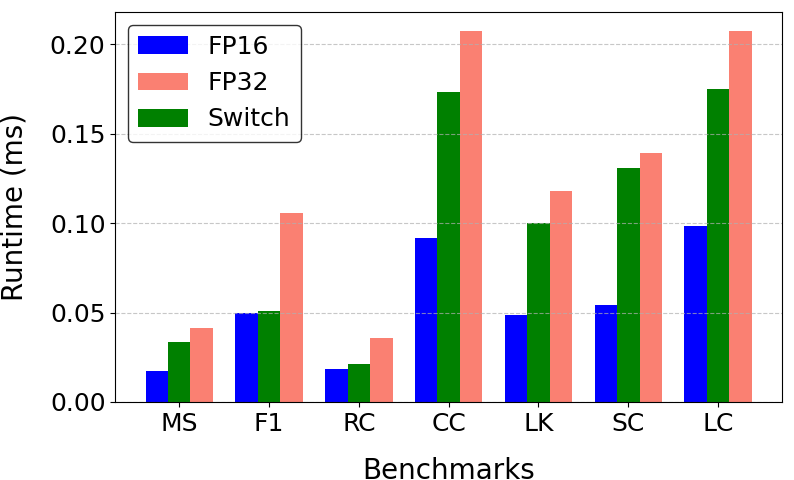}
    \caption{Switching schedule vs baselines in terms of runtime}
    \label{fig:switch_vs_FP16_vs_FP32_runtime}
\end{figure}

Next, we compare the runtime performance of our switching schedule against the FP16 and FP32 baselines. The reported runtime for the switching schedule includes the execution time of samples at different precisions as well as the switching overheads averaged over 3 runs. As shown in Figure~\ref{fig:switch_vs_FP16_vs_FP32_runtime}, the switching schedule achieves a runtime between FP32 and FP16, effectively striking a balance between computational cost and precision. For some systems, such as \texttt{F1} and \texttt{RC}, the runtime is close to that of FP16, primarily due to a higher proportion of FP16 samples and less switching overhead.

In summary, our optimal precision-switching schedule delivers near-FP32 control performance requirements while maintaining close-to-FP16 speed, achieving an effective trade-off between control quality and computational efficiency. These results highlight the importance of incorporating precision considerations into scheduling decisions and demonstrate that our approach offers a practical and efficient solution for realistic control systems.

%% file: sections/related.tex
\section{Related Work}
\label{sec:related}

\paragraph{Control Scheduling}
A substantial body of work has addressed control task scheduling from different perspectives, including energy efficiency, safety, and stability guaranties while scheduling a single controller on the processor. 
For example,~\cite{scheduling_single-energy} proposes energy-based scheduling by switching between multiple sampling periods at runtime.
In the context of skipped control execution or timing uncertainties, some recent works address aspects like stability~\cite{DMAC_vreman,maggio2020stability} and safety~\cite{hobbs2022safety} of the system while scheduling.

\paragraph{Control Performance Characteristics-Aware Scheduling.}
Several works have explored control performance characteristics, such as control cost, stability, safety, etc. in scheduling. Approaches such as ~\cite{maggio2020stability,ghosh_17,ghosh2020pattern,banerjee_P2SDS,banerjee_FMSS,vreman2021stability,online2018stability,ghosh_19}, synthesize schedules that preserve system stability even when control executions are occasionally delayed or skipped. Moreover, performance is assessed using metrics such as LQR cost~\cite{ghosh_17}, quadratic cost~\cite{vreman2021stability} or quadratic performance index~\cite{online2018stability} in the context of skipped control executions. We also use the LQR cost to measure control performance; however, we consider a strictly hard real-time context, where no deadlines are missed. 
Several works have addressed the co-scheduling of multiple control tasks on shared processors using methods such as response-time analysis~\cite{online2018stability}, earliest-deadline-first (EDF) scheduling~\cite{scheduling_codesign_stability,ghosh_17}, or SMT-based formulations~\cite{banerjee_FMSS}.
While our work currently focuses on a single control task, the proposed approach can be extended to multi-task co-scheduling in the future.

However, all scheduling techniques from the literature do not consider the impact of finite-precision on system's performance requirements ---an aspect we explicitly address in this work.

\paragraph{Finite-Precision Control and Safety Verification}
Several recent works have explored precision considerations in controller implementation and safety verification.
For instance,~\cite{precision2019memory} designs robust model-predictive controllers under finite precision, ensuring that deviations in the implemented controller remain within the robustness margin of the original design.
In contrast,~\cite{darulova2013synthesis,lohar2023sound} address the problem from the implementation side by generating sound and efficient implementations---either through heuristic search in a general framework or by formulating and solving a mixed-integer programming problem for neural network controllers.
More recent approaches~\cite{harikishan2024precision,teuber2025good} combine finite- and infinite-time safety guarantees with provably safe controller implementations.
While these works ensure safety under fixed-point or mixed-precision implementations, they do not consider \emph{when and how} to schedule different precision versions of a controller at runtime to balance control performance and computational efficiency. While we focus on floating-point precision, our approach can be easily adapted for fixed-point precision as well.

A recent work~\cite{idle_EDF_precision} shares a similar motivation by exploring precision-aware scheduling of multiple neural network–based real-time tasks. It proposes an idle-aware EDF policy to co-schedule tasks on CPU–FPGA platforms, allowing each task to choose among multiple precision levels to maximize accuracy within deadlines. However, unlike our approach, it does not address feedback controllers or consider control performance requirements.

%% file: sections/conclusion.tex
\section{Conclusion}
\label{sec:conclusion}
This paper introduces a novel precision-switching schedule framework that determines when a controller should switch between different floating-point precision levels to achieve near-optimal control performance with reduced computational cost while meeting the settling time requirement. We formulate the scheduling problem as a multi-objective MIQP problem, leveraging sound roundoff error bounds and linearized constraints to enable efficient solving with state-of-the-art optimizers. Experimental results on a suite of automotive control benchmarks show that our method efficiently generates optimized schedules between 32-bit and 16-bit floating-point precisions, achieving near–32-bit control performance at close-to–16-bit execution speed. These results confirm that our approach is an effective solution for precision-aware control scheduling and it lays the groundwork for multi-controller co-scheduling and support for various precision levels and data types.